\documentstyle[twocolumn,prd,aps]{revtex}

\input epsf

\begin{document}
\twocolumn[\hsize\textwidth\columnwidth\hsize\csname
@twocolumnfalse\endcsname

\title{Generic Tracking of Multiple Apparent Horizons with Level Flow}

\author{Deirdre M. Shoemaker$^1$, Mijan F. Huq$^1$, Richard A. Matzner$^2$}

\address{\small {$^1$Center for Gravitational Physics $\&$ Geometry, Department
of Astronomy $\&$ Astrophysics, The Pennsylvania
State University, PA 16802 \\  $^2$Center for Relativity, The University of Texas at Austin, 
Austin, TX 78712-1081}}

\maketitle

\begin{abstract}
We report the development of the first apparent horizon locator capable
of finding multiple apparent horizons in a ``generic'' numerical black hole
spacetime. We use a level-flow method which, starting from a single arbitrary
initial trial surface, can undergo topology changes as it flows towards
disjoint apparent horizons if they are present.  The level flow method has
two advantages: 1) The solution is independent of changes in the
initial guess and 2) The solution can have multiple components.
We illustrate our method of locating apparent horizons by tracking horizon
components in a short Kerr-Schild binary black hole grazing collision.

\end{abstract}

\pacs{PACS numbers: 04.70.Bw,04.25.Dm}

\vskip2pc]

\section{Introduction}
\label{sec:intro}
Our goal is to investigate the strong field regime of general relativity.
In particular we wish to focus on the study of coalescing black hole binaries.
Over the last three decades since the pioneering work of Cadez, Smarr, Eppley
and others, advances in computing technology, numerical algorithms and
techniques and our understanding of the underlying physics have advanced
to a point where we are able to carry out simulations of binary black hole
collisions in 3+1 dimensions. One of the outcomes of such simulations will be
an understanding of the underlying physics of the problem; and, therefore, a
prediction and understanding of the gravitational radiation content.
A detailed knowledge of how the resultant gravitational waveforms relate
to the physical parameters of the binaries that produce them will be of
importance to gravitational wave observatories (such as LIGO, VIRGO, TAMA,
GEO600) now under construction around the world.  
With the building of these  observatories, we stand at the epoch of
the first direct observations of astrophysical sources that involve strong 
field general relativity.

The orbit and merger of two black holes is one candidate source for ground 
based detection of gravitational waveforms.  This is of great interest to the 
relativity community.  The binary black hole problem is a two-body problem
in general relativity.  It is a stringent dynamical
test of the theory.  However, studying spacetimes
containing multiple black holes involves solving the Einstein equations,
a complex system of non-linear, dynamic,
elliptic-hyperbolic equations intractable in closed form.
The intractability of the problem has led to 
the development of numerical codes capable
of solving the Einstein equations.
 
Our approach to numerically solving the Einstein
equations involves reformulating them as an initial value problem.
In this  3+1 formulation \cite{York:ADM}, spacelike hypersurfaces 
parameterized in time foliate the spacetime.  The resulting
equations are coupled elliptic and hyperbolic differential
equations of the 3-metric, $g_{ij}$, and the extrinsic
curvature, $K_{ij}$.  The initial value problem
is solved by specifying a hypersurface at an instant
of time, say $t=0$, and evolving to the next hypersurface
at time $t=\delta t$ with the evolution equations 
to obtain $g_{ij}$ and $K_{ij}$ at the next time $t=\delta t$.

One vital issue in numerically solving the Einstein
equation describing spacetimes containing black holes is handling
the physical singularities.  As one approaches the singularity,
the values of the fields being computed approach infinity;
therefore, a region containing the singularity must
be avoided to keep the computation from halting.
Excision techniques are 
promising in avoiding the singularity. Excising
the singularity  involves locating a region interior to 
the event horizon containing the singularity on each 
evolving hypersurface.  This region is then ``masked''
from the computation.  The derivatives at the boundary
between the masked region and the computational domain
are handled using causal differencing, a differencing
scheme\cite{seidelsuen} that respects the causal structure of the spacetime.

In deciding where in space to excise we use the apparent horizon as opposed
to the event horizon.  By its very nature, the event horizon is a global 
construct depending on the entire spacetime.  The apparent horizon,
a local, {\it i.e.} spacelike 2-surface is more suitable.
Following a suggestion by Unruh \cite{Thornburg:1987}, 
the apparent horizon is used to define
the excised region to be masked at each time during the evolution.
Apparent horizons are defined locally for each time, and
exist at, or inside of, the event horizon.  In some spacetimes,
choices of foliation may lead to the absence of an apparent horizon. 
When the discussions in this refer to a hypersurface, it is
assumed an apparent horizon exists on that hypersurface.

Recent three-dimensional horizon locator codes
are capable of finding the location
of an apparent horizon
in generic {\it single} black hole spacetimes
\cite{Anninois,Baumgarte,Gundlach,HCM,Kemball,Nakamura}
and two \cite{Diener,Pasch}
are capable of finding disjoint multiple apparent horizons in 
the special case of conformally flat binary, time-symmetric 
black hole spacetimes.
Multiple apparent horizon finding algorithms will be necessary
in simulations of generic binary black hole spacetimes.  The method
presented in this paper, called the level flow method,
is capable of detecting multiple
apparent horizons in generic spacetimes.
The level flow algorithm 
has two advantages: 1) It is independent of a good initial guess
and 2) It is capable of following the surface through a 
change in topology.
In level flow, the 
apparent horizon equation is 
reformulated as a parabolic equation and a set of surfaces
are flowed with speeds dependent
on the expansion of the outgoing null vectors normal to each surface.

The purpose behind developing the level flow method of tracking
apparent horizons is to have a method capable of detecting
multiple apparent horizons on any given hypersurface 
without a good guess.  Specifically, we want a tracker 
that can detect the transition from a double to a single apparent
horizon in single time step without {\it a prior} knowledge of
the transition.   

In the rest of this paper we discuss apparent horizons in general
and current 3D
work in $\S~\ref{sec:ah}$ and  $\S~\ref{sec:method}$.
In  $\S~\ref{sec:LFM}$
we describe the level flow method in detail
and give a  brief description
of the numerical method involved in solving the apparent horizon
is in $\S~\ref{sec:NM}$.
We demonstrate the use of the level flow tracker on model
data and a binary black hole grazing collision in
$\S~\ref{sec:testing}$ and $\S~\ref{sec:graze}$.

\section{Apparent Horizons}
\label{sec:ah}

\begin{figure}[h]
\epsfxsize=6cm
\centerline{\epsfbox{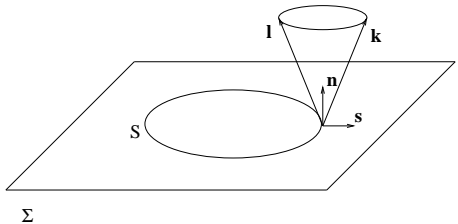}}
\vspace{0.5cm}
\caption{Representation of a 2-sphere embedded in a hypersurface, $\Sigma$.}
\label{fig:aheqn}
\end{figure}

$M$ is the spacetime with metric ${}^4g_{ab}$ foliated
by hypersurfaces $\Sigma$ parameterized by $t$ with 3-metric $g_{ab}$.
Let $S$ be a surface with $S^2$ topology on $\Sigma$.  The apparent
horizon is the outermost marginally trapped surface in $\Sigma$, a
surface with zero expansion.
The zero expansion of the surface is defined in terms of
outgoing null vectors to $S$, $k^a$, such that $k^a$
have zero divergence
\begin{equation}
\nabla_a k^a = 0,
\end{equation}
where $\nabla_a$ is the covariant derivative associated with ${}^4g_{ab}$.
Referring to fig.~(\ref{fig:aheqn}), $k^a$ is defined
in terms of the spacelike normals to $S$, $s^a$, and the
future directed timelike normals, $n^a$, such that:
\begin{equation}
k^a = s^a + n^a.
\label{eqn:ksn}
\end{equation}
The expansion of the outgoing normals, $\nabla_a k^a$,
can be rewritten in terms of quantities defined on the hypersurface:
\begin{equation}
\kappa \equiv D^a s_a - K + s^a s^b K_{ab}.
\label{eqn:kappa}
\end{equation}
$D_a$ denotes covariant differentiation with respect to $g_{ab}$,
$K_{ab}$ is the extrinsic curvature, and $K$ is $g^{ab}K_{ab}$.
In fact, there is a level set of surfaces in $\Sigma$ parameterized 
by $\kappa$.
Each surface in the level set is defined by the constant expansion
of its null vectors such that 
\begin{equation}
\kappa = c_n,
\label{eqn:cn}
\end{equation}
where $c_n$ are constants labeled by the positive integer $n$.
Marginally trapped surfaces are members of this set for
\begin{equation}
\kappa = 0.
\label{eqn:aheqn}
\end{equation}
Eqn.(\ref{eqn:aheqn}) is called the apparent horizon equation since
the apparent horizon is the outermost surface with  $\kappa = 0$ 
in $\Sigma$.

The $S^2$ topology of the apparent horizon naturally lends itself
to characterization via spherical coordinates.  The function,
\begin{equation}
\psi = r - h(\theta,\phi)
\label{eqn:shapefunc}
\end{equation}
is a level set of 2-spheres in $\Sigma$,
where $h(\theta,\phi)$ 
is called the apparent horizon shape
function.  A
marginally trapped surface has $\psi = 0$.
The spacelike normals to $S$ are defined from 
eqn.~(\ref{eqn:shapefunc}) such that
\begin{equation}
s^i = g^{ij} \partial_j \psi / \sqrt{g^{kl}\partial_k \psi \partial_l \psi}
\label{eqn:spacenorms}
\end{equation}
is the spacelike vector field at every point of $S$.
The apparent horizon equation in spherical coordinates ($h(\theta,\phi),\theta,\phi$) is
a 2-dimensional problem in $\theta$ and $\phi$.

\section{Current 3-D Methods}
\label{sec:method}

The approaches to solving the apparent horizon equation on a 
three-dimensional hypersurface can be addressed roughly in two 
categories: methods that solve the apparent horizon equation directly
and methods that solve it by first recasting it as a parabolic equation.
This paper does not address spherical and axi-symmetric approaches.
One of the first three-dimensional apparent horizon solvers was
published by Nakamura, Kojima, and Oohara \cite{Nakamura}.  They directly
solve the apparent horizon equation by using spherical harmonic
basis functions to expand the apparent horizon shape function, 
$h(\theta,\phi)$ into 
\begin{equation}
h(\theta,\phi) =
 \sum^{l_{max}}_{l=0} \sum^{l}_{m=-l} a_{lm}Y_{lm}(\theta,\phi).
\end{equation}
This method is called the pseudospectral method.
A finite number of the coefficients, $\{a_{lm}\}$ parameterize
the horizon shape function, and the maximum $l_{max}$ depends
on the computation and deviations from a sphere.
The apparent horizon equation can then be solved by writing
it as
\begin{equation}
\|\kappa(a_{lm})\| = 0,
\end{equation}
and using functional integration routines to find the 
coefficients $a_{lm}$.  
Others have used similar methods \cite{Baumgarte,Kemball,Anninois}.

In another approach to direct solutions of the apparent horizon equation
is to treat it as a boundary value problem. One notes that a
discretization
of this equation leads to a system of algebraic equations which can then
be solved via Newton's method. Thornburg \cite{Thornburg}
discusses applications of
Newton's method to this problem in general and shows results in
axisymmetry.
Huq \cite{Huq} has implemented a similar algorithm based on Newton's method that
utilizes Cartesian coordinates to difference the equations.

Tod \cite{Tod} first suggested the use of curvature flow in 
solving the apparent horizon equation by recasting it as a parabolic
equation.  Bernstein \cite{Bernstein} implemented Tod's suggestion in 
axisymmetry.  Gundlach \cite{Gundlach} introduced 
a fast flow method which combines the 
ideas of the flow method with the pseudospectral method.
Pasch \cite{Pasch} and Diener \cite {Diener}
implemented a similar method, a level-set method, in three-dimensions 
and found discrete  apparent horizons in multi-black-hole spacetimes;
however these spacetimes were confined to be conformally flat and time-symmetric.

Each of the approaches briefly described above, solving
the apparent horizon equation directly or solving it {\it via} a 
parabolic equation, has its advantages.  Direct methods tend to 
be faster while flow methods do not rely on ``good'' initial guesses.
However, none of these methods are applicable to the generic, 
multi-black-hole problem.  Herein lies the motivation behind the level flow
method.   The level flow method is the only method designed to locate discrete apparent
horizons in generic spacetimes containing one or two discrete horizons.   

\section{Level Flow Method}
\label{sec:LFM}
\subsection{Curvature Flow}
The level flow method is a hybrid flow$/$level$-$set
method.  The previous section mentioned the 
flow method, this section gives more detail on the flow
method which is 
the foundation of the level flow method.
The flow approach, as suggested by Tod, is to rewrite the apparent
horizon equation as the speed function in a parabolic equation.
In the case of a time symmetric hypersurface, in which $K_{ab} = 0$,
the apparent horizon equation reduces to the condition for a minimal surface,
$D_a s^a =0$.  In this case, the surface $S$ is at a local 
extremum of the area.
The starting surface,
$S(\lambda=0)$, is parameterized by coordinates $x^a$
and evolved in terms of a parameter $\lambda$.
The equation of motion is
\begin{equation}
\frac{\partial x^a}{\partial \lambda} = - H s^a
\label{eqn:mcf}
\end{equation}
where $\partial x^a/\partial \lambda$
is a vector field, and  $H$  is the mean
curvature, which is the trace of extrinsic curvature
associated with embedding $S$ in $\Sigma$ given by
\begin{equation}
H = D_a s^a.
\label{eqn:mean}
\end{equation}
Eqn.(\ref{eqn:mcf}) is the gradient flow for the area
functional.  The area decreases monotonically with increasing $\lambda$.
Grayson \cite{Grayson}  has
shown that a surface deforming under its gradient field
(Eqn.(\ref{eqn:mcf})) will evolve to a stable
minimum surface (surface is local minimum of area)
if there is one, otherwise to a point.

In numerical relativity, we are interested
in the generic case, with $K_{ab} \neq 0$, for
which the marginally trapped surfaces
differ from minimal surfaces, 
the surfaces are not extrema of the area.
However, Tod suggests an equation similar to  Eqn.(\ref{eqn:mcf})
as a curvature flow:
\begin{equation}
\frac{\partial x^a}{\partial \lambda} =  F(\kappa) s^a
\label{eqn:flow}
\end{equation}
using $F(\kappa) = -\kappa$ where $\kappa = D_as^a + s^as^bK_{ab} -K$ as in eqn.(\ref{eqn:kappa}).
We have found eqn.~(\ref{eqn:flow}) to be a successful practical
implementation of the flow method.

\subsection{Level Flow}
\label{sec:LSM}

Eqn.~(\ref{eqn:flow}) gives us an initial value problem.  Given
information about the system at some initial $\lambda$, eqn.~(\ref{eqn:flow})
will describe the system for all its future propagation in $\lambda$.
Directly solving eqn.~(\ref{eqn:flow}) will lead to a successful
detection of single apparent horizons; however, solving it directly
does not ensure correct handling of a topology change which is 
necessary for detection of multiple apparent horizons.
By combining the flow method with a level-set 
idea however, this topology change can be effected and multiple apparent 
horizons can be tracked starting from a single guess surface. 

First eqn.~(\ref{eqn:flow}) is recast from an equation governing
the motion of the coordinates parameterizing $S$, namely $x^a$, to
an equation governing the motion of the surface $\psi$.  
Noting that $\psi$ is parameterized by $\lambda$, 
\begin{eqnarray}
\frac{\partial \psi}{\partial \lambda} =
\frac{\partial x^a}{\partial \lambda}\frac{\partial \psi}{\partial x^a} 
\end{eqnarray}
by the chain rule, and multiplying eqn.(\ref{eqn:flow}) by $\frac{\partial \psi}
{\partial x^a}$ gives
\begin{equation}
\frac{\partial \psi}{\partial \lambda} =
F(\kappa) s^a \partial \psi/\partial x^a.
\end{equation}
Using
\begin{equation}
s^a = g^{ab}\frac{\partial \psi}
{\partial x^b}/\| \nabla \psi\|
\end{equation}
and
\begin{equation}
\|\nabla \psi\| = \sqrt{\frac{\partial \psi}{\partial x^a} \frac
{\partial \psi}{\partial x^b} g^{ab}},
\label{eqn:nablapsi}
\end{equation}
the test surface's flow is given by:
\begin{equation}
\frac{\partial \psi}{\partial \lambda}  =  F(\kappa) \|\nabla \psi\|.
\label{eqn:flow0}
\end{equation}
Eqn.(\ref{eqn:flow0}) is a reformulation of eqn.~(\ref{eqn:flow}) and will
flow the surface, $\psi$, to a marginally trapped surface at $\psi=0$ when 
$\kappa=0$.

\begin{figure}[h]
\vspace{0.5cm}
\epsfxsize=4cm \epsfysize=7cm
\centerline{\epsfbox{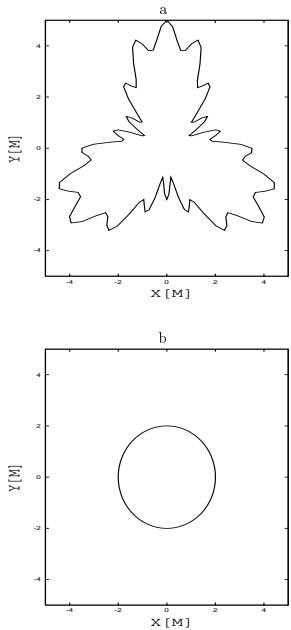}}
\vspace{0.5cm}
\caption[With a flow method, an arbitrary initial guess still flows to 
the solution.]
{Example of a surface undergoing flow on a spherically symmetric hypersurface.
The figure labeled (a) is $\psi(\lambda=0)$, the initial guess.  Figure (b) is
$\psi(\lambda)$ representing the final solution, the apparent horizon.}  
\label{fig:leaf}
\end{figure}

The strength of the flow methods is their ability
to locate a surface with $\kappa =0$ given any non-pathological initial
surface.  For example, the apparent horizon in a spherically symmetric
spacetime (Schwarzschild) was found by flowing an initial surface shaped
as a leaf, see Fig.~(\ref{fig:leaf}).  This ability
is especially important when tracking horizons during evolutions
of binary black hole spacetimes.  In this case, finding apparent horizons
for two discrete apparent horizons in each $\Sigma(t)$ involves flowing
two initial guesses, one for each horizon \cite{Bruegmann}.
On $\Sigma(t=0)$, the location
of the apparent horizons may be known; however, as the black holes accelerate 
the task of guessing the locations of the two horizons becomes more difficult.
Further, the two horizons merge into a single horizon at a single instant of
time, rendering a good initial guess difficult.  Some way of determining
when two horizons merge into a single horizon is necessary.  The level flow
method takes care of these issues by not requiring a good initial guess 
$(\psi (\lambda=0))$ and by detecting multiple apparent horizons from a single
guess $(\psi (\lambda=0))$. 
\vskip \baselineskip

The level flow method differs from the flow method
in the specification of the speed function, $F(\kappa)$.
$F(\kappa)=0$ determines when the propagation of the
surface stops. The flow method is in the form of eqn.~(\ref{eqn:flow}),
in which $F(\kappa) = -\kappa$.   A
good choice since $F(\kappa)=-\kappa=0$ indicates
a marginally trapped surface; but this choice
will not flow $\psi$ though a fission.  In general
the scheme fails as the surface pinches due to ill-defined normals at the surface.
The level flow method alleviates this problem by looking for indications
that the surface topology is about to change before the pinching occurs.  
(Another method which was introduced by Sethian and Osher \cite{Sethian}
for non-relativistic problems is to flow a higher dimensional surface in which 
$\psi$ is embedded.  This higher dimensional surface does not fission.
This has only been implemented in a time-symmetric spacetime \cite{Pasch}
and requires more computational power due to the extra dimension in the problem.) 

The level flow method flows a set of two-dimensional surfaces in $\Sigma$
parameterized by $\kappa$.  
We call the set of surfaces a level set and label the set $S(c_n)$.
Each surface has a constant value of $\kappa=c_n$ everywhere on it.
The set of surfaces is defined by varying $c_n$ as the flow progresses
\begin{equation}
c_{n+1} = c_n \pm \Delta c,
\label{eqn:c}
\end{equation}
where $(+)$ indicates outward flow, $(-)$ inward flow, and $\Delta c \propto
\parallel \kappa \parallel_2$.
Each surface obeys the equation of motion given in  eqn.~(\ref{eqn:flow0})
with $F(\kappa)$ defined to flow to multiple surfaces.
We choose two
options for the speed function:
\begin{eqnarray}
F(\kappa) &=& \kappa -c_n \\
F(\kappa) &=&  (\kappa -c_n) \arctan^2(\frac{\kappa-c_n}{\kappa_o}).
\label{eqn:atan}
\end{eqnarray}
As $\kappa-c_n \rightarrow 0$, both functions are
solving for a particular surface in the level set, $S(c_n)$.
The second function, eqn.(\ref{eqn:atan}),
behaves similarly to the first but allows
for larger time steps near
a fissioning surface because it moves points further from the
$\kappa - c_n=0$ surface faster than the points closer to this surface.

Eqn.~(\ref{eqn:flow0}) is an initial value problem 
requiring $\psi$ to be specified
at $\lambda = 0$. 
To initialize the starting surface, we need only supply
an origin and radius.  Taking into account that there may be more
than one marginally trapped surface, it is best to start with
an initial surface larger than the expected horizon.
The values of $g_{ij}$ and $K_{ij}$ are required everywhere
on the surface to evaluate $\kappa$.  These functions can be
known analytically or generated by evolution codes. 
As the flow velocity approaches zero, $F(\kappa-c)\rightarrow 0$,
$\kappa \sim c_n$ and a surface $S(c_n)$ is found
within a tolerance ($\epsilon_{\kappa}$). When
$\kappa = 0$,
the located surface describes a
marginally trapped surface.
\begin{figure}[h]
\epsfxsize=4cm
\centerline{\epsfbox{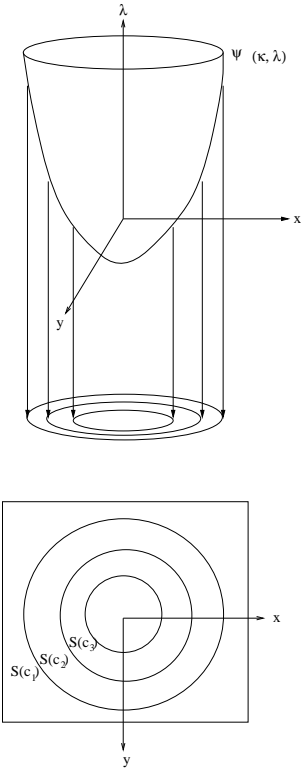}}
\vspace{0.5cm}
\caption[The level flow method uses a set of surfaces.]
{Schematic of $\psi(c_n)$, the level set of surfaces
in $\Sigma(t)$.
We solve for a single surface, $S(c_n)$, in $\psi(c_n)$.
Multiple levels are used in detecting the existence of multiple horizons.}
\label{fig:proj}
\end{figure}

\begin{figure}[h]
\epsfxsize=7cm
\centerline{\epsfbox{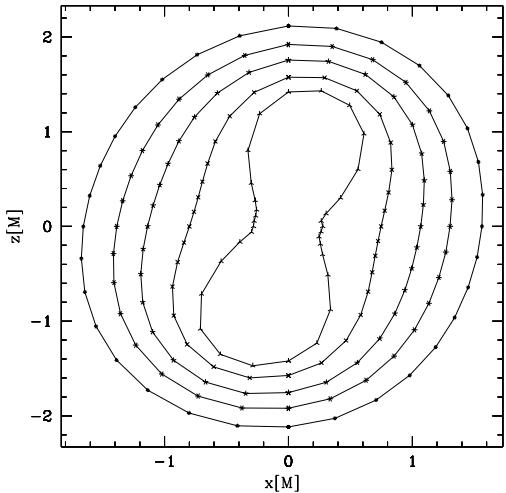}}
\vspace{0.5cm}
\caption{Plot of five levels ($\kappa$ = 0.14,0.12,0.10,0.08,0.06) of constant
divergence of outgoing
null geodesics.}
\label{fig:level}
\end{figure}

Fig.(\ref{fig:level}) shows
the level set found in a spacetime containing
two black holes with coordinate
locations $(-0.954,0,-0.3)M$ and $(0.954,0,0.3)M$.
Each 2-surface has a constant value
of $\kappa$.  We monitor the topology of the deforming surface
by computing the radial component of the gradient of
$\kappa$ with respect to the normals of
each surface in the level set.  The gradient is defined as:
\begin{equation}
\frac{\|\kappa_{n-1} - \kappa_{n}\|}{\|\psi_{n-1} - \psi_{n}\|},
\end{equation}
where $\psi$ is the function given in eqn.(\ref{eqn:shapefunc}).
A sharp increase in the gradient indicates the
existence of multiple surfaces.
To ensure that we do not erroneously abandon
a single surface, we also monitor the maximum of the $l_2$-norm of $\kappa$. 
If $\kappa$ is no longer decreasing,
we are no longer finding a solution
to eqn.(\ref{eqn:cn}); otherwise the single surface is retained.
The level flow method
is essentially a special set of surfaces with properties
that let us determine when to break.  If we only
flowed to $\kappa = 0$, we would not form
the collection of $\kappa=$constant surfaces.

\begin{figure}[h]
\epsfxsize=7cm
\centerline{\epsfbox{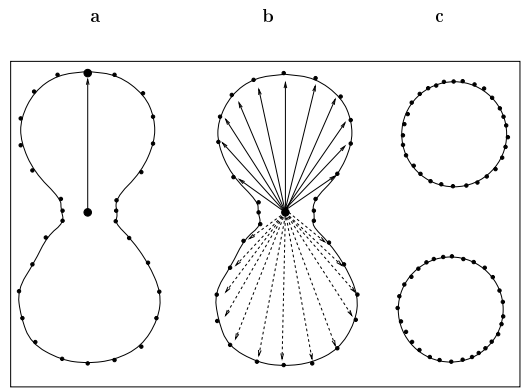}}
\vspace{0.5cm}
\caption[Representation of the decision making process for multiple
surfaces in the level flow method]
{Two-dimensional schematic representation of the three-dimensional
decision process to identify the two surfaces
that will evolve to the two apparent horizons.  Fig.~(c)
depicts the two surfaces that will act
as new test surfaces. }
\label{fig:yikes}
\end{figure}

Once a topology change is indicated, the radii and origins for
each of  the new surfaces
are found (note that these four parameters for each surface
are all that is  needed to initiate two new trial surfaces).
These origins
and radii are determined using the 
location of the last of the single surfaces.
Using this last single surface, we can find the origin 
of the last surface and  the location on the surface
with minimum gradient of the value of $\kappa$.
This occurs at
the farthest points from the pinching in the surface.
Picture (a) in fig.~(\ref{fig:yikes}) 
shows the last single surface with an arrow drawn from
the  origin to the 
point on the surface with a minimal change in $\kappa$. 
The arrow indicates a chosen direction.  All points lying
in this direction are collected and averaged
to find a radius and center of mass.  All points lying in the
opposite direction are also collected and used
to calculate the radius and center of mass for the second surface
using the dotted arrows in picture (b) of fig.(\ref{fig:yikes}).
These two sets of radii and centers are the initial starting
parameters for each new trial surfaces.  
The tracker then flows the two new surfaces depicted
in (c) of fig.(\ref{fig:yikes}) until
$\kappa = 0$ within $\epsilon_{\kappa}$.
The level flow code is only started by the user once, the
subsequent flowing to multiple apparent horizons is done automatically.

The advantage to the level flow method is its capability
to detect apparent horizons in generic, multiple black hole spacetimes 
from a single reasonable initial guess.
The drawbacks are the dependence
of $\Delta \lambda$
on the spatial grid size,
$\Delta \lambda \sim N^{-2}$ where $N^2=N_{\theta}N_{\phi}$
is the number of
grid points, and the fact that we flow to a speed of zero (the
flow speed approaches zero as $\kappa$ approaches zero).
When using apparent horizon tracking in our evolution code,
we will not require knowledge of the apparent horizon location to
high precision; in fact we can find a surface with $\kappa \le 0$
to remove the singularity thus alleviating some of the speed issues.   
Nonetheless, we plan to 
improve the speed of this algorithm. 
Some improvements have already been made to increase the efficiency
of the current algorithm.  The addition of the $\arctan^2$ function,
eqn.(\ref{eqn:atan}), speeds up the algorithm during the fissioning
process.  Monitoring the number of steps needed to complete
a Crank-Nicholson iteration (see $\S$\ref{sec:NM}))
has also proven useful in increasing
efficiency.

\section{Numerical Method}
\label{sec:NM}
The previous section described how the level flow method
is used to solve the apparent horizon equation.  The resulting
parabolic equation is updated using an iterative Crank-Nicholson
method updating the variables at every $\lambda$-step.  
Iterative Crank-Nicholson converges to an exact
solution of the implicit problem.  However, the detailed behavior
of this convergence \cite{Teukolsky} shows that the Crank-Nicholson solution
at a particular iteration has an amplification factor
$|{\cal{A}}^{(n)}|$
that oscillates around unity.  The behavior varies in pairs:
$|{\cal{A}}^{(n)}| < 1$ for $n = 2,3$;
$|{\cal{A}}^{(n)}| > 1$ for $n = 4,5$, etc.
while $|\hspace{.1cm}|{\cal{A}}^{(n)}|-1| \rightarrow 0$ monotonically as
$n \rightarrow \infty$.  $n$ is counting
the number of iterations it takes to get $\hat{\kappa} = \kappa$ within
the specified tolerance.
For the data presented here, a Crank-Nicholson iteration of $n=2$ or $n=3$ was maintained
for errors less than the spatial grid spacing squared, $h^2$.

\begin{figure}[h]
\epsfxsize=6cm
\centerline{\epsfbox{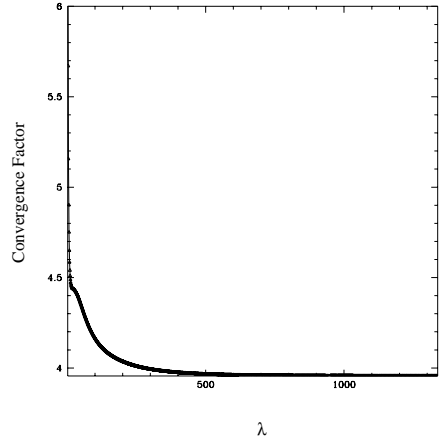}}
\vspace{0.5cm}
\caption[The convergence factor for $\kappa$ in Schwarzschild data]
{Convergence factor for radial variable $\kappa$
with $h=0.05$, $2h=0.10$, $4h=0.20$, and $\Delta \lambda = 0.0012$ for Schwarzschild data.
Second order accuracy is obtained.}
\label{fig:cnvg2}
\end{figure}

The spatial derivatives are 
approximated to second order in truncation error using
centered finite differencing molecules.
To verify the convergence of the level flow code, 
we include a plot of the convergence factor versus the number
of iterations taken in fig.~(\ref{fig:cnvg2}).  
The convergence factor is given by
\begin{equation}
C_f \equiv \frac{{\hat{\kappa}}_{2h} - {\hat{\kappa}}_{4h}}{{\hat{\kappa}}_h
- {\hat{\kappa}}_{2h}},
\label{eqn:convergence}
\end{equation}
where ${\hat{\kappa}}$ is the discretized $\kappa$ and $h$ is the spatial grid spacing.  
For a second order scheme, the convergence factor in
eqn.~(\ref{eqn:convergence}) is $C_f = 4 + O(h^2)$.

For the closed form solutions detailed in the next section,
the data are given by evaluating the closed form analytically on
the two-surface.  However, the goal is to use the level flow method
during an evolution
including evolutions involving a region excised from computational
consideration.  
The approach we take to evaluate $\kappa$ from a Cartesian
grid of data $(g_{ij}, K_{ij})$ is the same as that used and
described in Huq \cite{Huq}. This approach discretizes the apparent horizon
equation using Cartesian coordinates on 3d-stencils centered on points
on the surface. These stencils are not aligned with the 3d-lattice from
which we obtain $g_{ij}$ and $K_{ij}$ data. Our apparent horizon surfaces
are embedded in such lattices and as a result interpolations
must be carried out to obtain the metric data on the surface as it
evolves.
The algorithms and methodology for evaluating $\kappa$ are described in
detail in \cite{HCM,Huq}. 

\section{Testing the Method with Closed Form Solutions}
\label{sec:testing}
The level flow method of tracking apparent horizons has been designed to
locate apparent horizons in single and multiple black hole spacetimes.
To test the level flow tracker, we locate apparent
horizons in Schwarzschild, Kerr,
and Brill-Lindquist data.
In particular, we also demonstrate the level flow method's ability
to detect binary black holes in the Brill-Lindquist data.

\subsection{Schwarzschild Data}
\label{sec:schwarz}

The Kerr-Schild metric provides a closed-form
description of both the Schwarzschild
and the Kerr solutions to the Einstein equation and is given by:
\begin{equation}
g_{ab} = \eta_{ab} + 2Hl_{a} l_{b},
\label{eqn:KSmetric}
\end{equation}
where $\eta_{ab}$ is the Minkowski metric,
$\eta_{ab}=$diag$(-1,1,1,1)$.  $H$ is a scalar function of the coordinates
and $l_{a}$ is an ingoing null vector with respect to both
the Minkowski and full metrics; that is
$l_a$ satisfies the relation:
\begin{equation}
\eta^{ab}l_{a}l_{b}=g^{ab}l_{a}l_{b}=0.
\end{equation}

For the ingoing Eddington-Finkelstein form
of the Schwarzschild solution, the metric given in eqn.(\ref{eqn:KSmetric})
has the scalar function, $H$,  given by:
\begin{equation}
H=\frac{M}{r}
\end{equation}
and the components of the null vector:
\begin{eqnarray}
l_t & = & 1 \\
l_x & = & \frac{x}{r} \\
l_y & = & \frac{y}{r} \\
l_z & = & \frac{z}{r}
\end{eqnarray}
where we have adopted rectangular coordinates ($t,x,y,z$) with $r=\sqrt{x^2+y^2+z^2}$,
and $M$ the mass of the black hole.

We track the apparent horizon in this situation for
a single black hole of mass $M$.  
The area and radius of the event horizon for the
Schwarzschild solution of the
Kerr-Schild metric is known in closed form \cite{Misner}
given by:
\begin{equation}
A = 4\pi r^2_+
\label{eqn:area}
\end{equation}
where $r_+$ is the event horizon radius and equals $2M$.
In the slicing we have chosen, the apparent horizon
coincides with the event horizon.
Using the level flow method we found the apparent horizon 
to converge to the closed form solution
giving a $0.35\%$ relative error at a course resolution 
($17 \times 17$ grid).  
The area of the tracked apparent horizon
is computed by
\begin{equation}
A_{num} \equiv \int_S \sqrt{h} dx dy,
\label{eqn:areaS2}
\end{equation}
and converges to the closed form solution, 
eqn.(\ref{eqn:area}).
In eqn.(\ref{eqn:areaS2}) $h$ is the determinant of the 2-metric $h_{ab}$
on the apparent horizon surface,
and $x$ and $y$ are surface coordinates.  The numerical area
is determined from eqn.(\ref{eqn:areaS2}) by
calculating the determinant at every point in the
grid and using a trapezoidal integration scheme \cite{Huq}.

\subsection{Kerr Data}

\begin{figure}[h]
\epsfxsize=9cm
\centerline{ \epsfbox{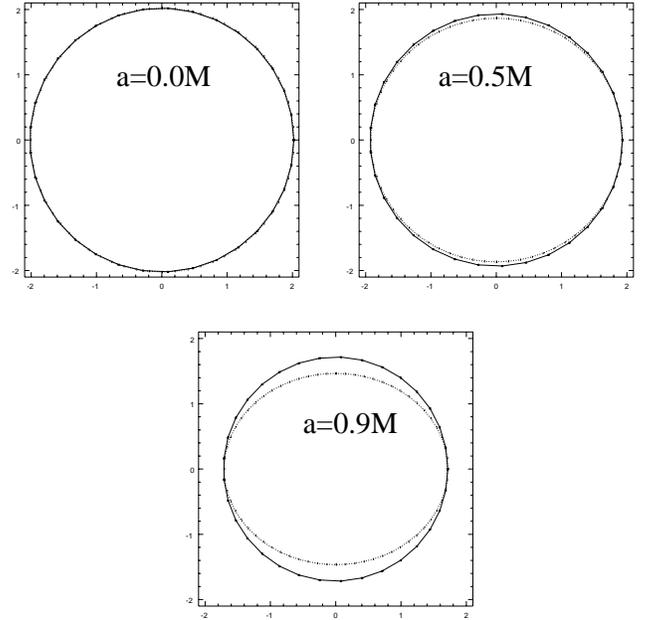}}
\caption
{The three plots correspond to the location
of the apparent horizons for black holes with three different
values of angular momentum.  The units of the graph are $M$,
the solid line is the $\theta = \pi/2$
slice and the dashed line is the $\phi = \pi$ slice.}
\label{fig:a}
\end{figure}

The Kerr solution is a second solution given by the
Kerr-Schild metric, eqn.(\ref{eqn:KSmetric}).
The Kerr solution is the solution for a spinning black hole,
{\it i.e.} a black hole with an internal  angular momentum per unit mass
given by $a$.
In rectangular coordinates $(t,x,y,z)$, the scalar function
and null vector are given by:
\begin{equation}
H = \frac{Mr^3}{r^2+a^2z^2}
\label{eqn:KSH}
\end{equation}
and
\begin{equation}
l_{\mu} = (1,\frac{rx+ay}{r^2+a^2},\frac{ry-ax}{r^2+a^2},\frac{z}{r}),
\label{eqn:KSl}
\end{equation}
where $\mu = (t,x,y,z)$, $M$ is the mass of the black
hole, $a=J/M$ is the angular momentum per unit mass
of the black hole in the z-direction, and r is obtained from:
\begin{equation}
\frac{x^2+y^2}{r^2+a^2}+ \frac{z^2}{r^2} = 1 :
\end{equation}
\begin{equation}
r^2 = \frac{1}{2}(\rho^2-a^2) + \sqrt{\frac{1}{4}(\rho^2-a^2)^2+a^2z^2},
\label{eqn:r}
\end{equation}
with $\rho = \sqrt{x^2+y^2+z^2}$.
The difference here is the addition of angular momentum.
We test two cases, $a=0.5M$ and $a=0.9M$.  Fig.(\ref{fig:a})
presents a Schwarzschild ($a=0M$) case, together with
the $a=0.5M$ and $a=0.9M$ cases.
The solid line is the $\theta = \pi/2$
slice and the dashed line is the $\phi = \pi$ slice.
We find the expected result, that the deformation in the $\phi = \pi$
slice increases with $a$.
The radius of the event horizon is given by
\begin{equation}
r_+ = M + \sqrt{M^2 - a^2}.
\label{eqn:eventra}
\end{equation}
The solution to eqn.(\ref{eqn:eventra}) for $a=0.5M$ is
$r_+=1.87M$ and the numerical solution we obtain for the horizon
radius is $r_{num}
= 1.87M$, with an error of $0.06\%$.
In the $a=0.9M$ case, $r_+=1.44M$ and $r_{num}= 1.46M$,
with a $1.39\%$ error.   The area of the horizon for each case
can be calculated using
\begin{equation}
A = 4 \pi (r_+^2 + a^2)
\label{eqn:areaa}
\end{equation}
(generalizing eqn.(\ref{eqn:area})), and compared to a numerical eqn.(\ref{eqn:areaS2}).
computation using eqn.(\ref{eqn:areaS2}). 
In the $a=0.5M$ case, eqn.(\ref{eqn:area}) gives $A = 46.89M^2$,
numerically we obtain $A_{num} = 46.88M^2$, resulting
in a $0.21\%$ error.  In the $a=0.9M$ case,
$A= 36.09M^2$ and $A_{num} = 36.39M^2$ with
a $0.83\%$ error.
The errors will decrease
as $\kappa$ is driven closer to zero.

\subsection{Brill-Lindquist Data}
\label{sec:bl}

In this section, we study a binary black hole system
using Brill-Lindquist data [\ref{Brill}].
These data are useful to us for two reasons:
We can verify previous results of the critical
separation, and study an
example of how the tracker works
in finding multiple apparent horizons.
The 3-metric is time symmetric, $K_{ab} = 0$,
and  is conformally flat:
\begin{equation}
g_{ab} = \phi^4 \eta_{ab}
\end{equation}
where
\begin{equation}
\phi = 1 + \sum^N_{i=1}\frac{M_i}{2r_i}
\end{equation}
and $N$ is the number of holes (here $N=2$),
$M_i$ is the mass of the ith black
hole
and the $r_i$ are the radial distances from 
the centers of the
black holes

We use isotropic coordinates to express the metric as
\begin{equation}
ds^2 = \phi^4 (dr^2 + r^2 d\theta^2 +  r^2 \sin^2\theta d\phi^2)
\end{equation}
with
\begin{equation}
r_i = \sqrt{r^2+d_i^2 - 2d_i r \cos\theta},
\end{equation}
where $d_i$ is the distance between the holes and the
center of the coordinate system.
When they are far apart, each hole has an apparent horizon of
radius $M/2$ in these coordinates.  The area of each of
the holes when they are well separated is $16\pi M^2$.

\begin{figure}[h]
\epsfxsize=7cm
\centerline{\epsfbox{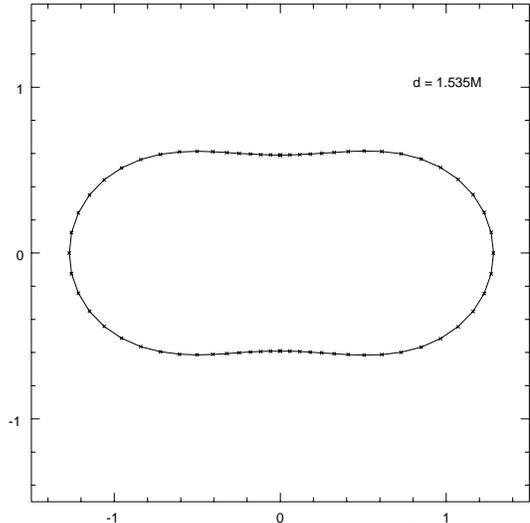}}
\caption[The apparent horizon for Brill-Lindquist data at
critical separation]{Separation of $1.535M$ with a $33^2$ grid. The area
was determined numerically to be $184.16M^2$}
\label{fig:d1.535}
\end{figure}

The limiting separation of the holes between single and double
horizons was found by Brill and Lindquist \cite{Brill} to be $1.56M$,
Cadez $1.534M\pm0.002M$~\cite{Cadez},
$[1.5M,1.6M]$ by Alcubierre {\it et al.}~\cite{Alcubierre}, and $1.535M$
by Huq \cite{Huq}.  We found
a critical separation 1.53(5)M.  The apparent horizon at
the critical separation of 1.535M is shown in fig.(\ref{fig:d1.535})
using the level flow code with $33^2$ grid points.

\begin{figure}[h]
\epsfxsize=7cm
\centerline{\epsfbox{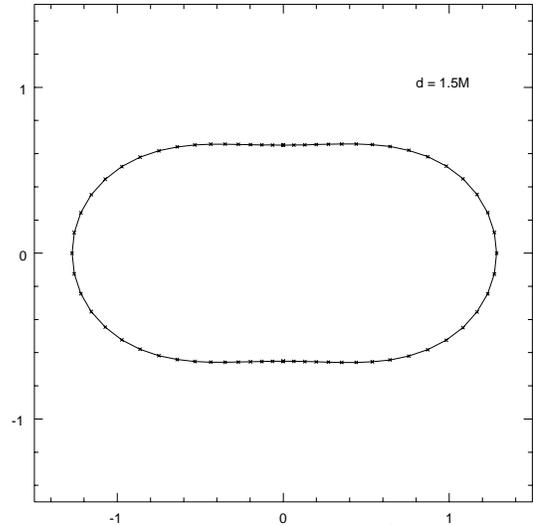}}
\caption[Brill-Lindquist data at a separation $1.5M$]
{Separation of $1.5M$ with a $33^2$ grid.  The
area is $185.41M^2$. }
\label{fig:d1.5}
\end{figure}

\begin{figure}[h]
\epsfxsize=7cm
\centerline{\epsfbox{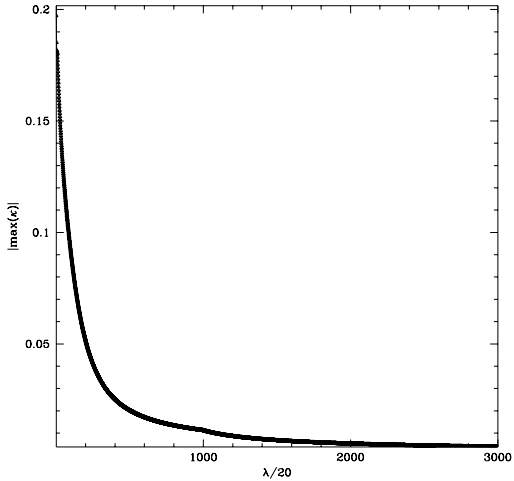}}
\caption[The absolute value of the maximum of $\kappa$ per iteration]
{The absolute value of the maximum of expansion, $\kappa$,
 per iteration,
$\lambda$ every 20th step.  The kinks at $\lambda = 1000$ and
$\lambda = 2000$ are from restarting the code with a different $\lambda$-step.}
\label{fig:absy1.5}
\end{figure}

\begin{figure}[h]
\epsfxsize=7cm
\centerline{\epsfbox{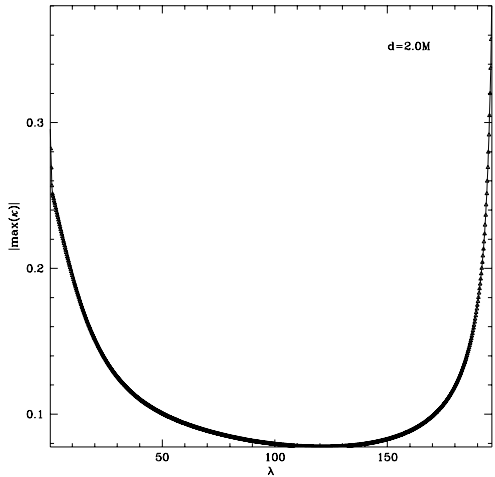}}
\caption[The absolute value of the maximum of expansion, $\kappa$,
 per iteration]
{The absolute value of the maximum of expansion, $\kappa$,
 per iteration is plotted.  The increase in the expansion is caused by
imminent fission.}
\label{fig:absy2.0}
\end{figure}

The horizon found for a separation of $d=1.5M$ which is less than the critical
separation, is shown in fig.(\ref{fig:d1.5}).
Fig.(\ref{fig:absy1.5}) is a plot of the $l_2$-norm of the maximum of
$\kappa(\theta)$
for the separation $d=1.5M$
at each iteration plotted versus the number of $\lambda$-steps.
This is one of the checks in the level flow code to
ensure that the apparent horizon equation is still being
solved.  We expect the expansion to continue to decrease
if we have started outside the apparent horizon and are flowing
inward.  As we will see in fig.(\ref{fig:absy2.0}), the
expansion increases as fission occurs in a data set with separated holes.

\begin{figure}[h]
\epsfxsize=6cm \epsfysize=7cm
\centerline{\epsfbox{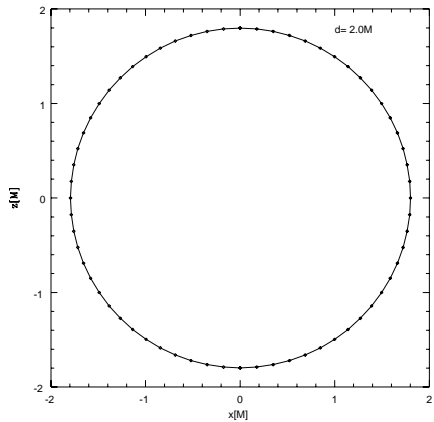}}
\caption{The starting surface of the level flow method for a
separation of d=2.0M}
\label{fig:initial}
\end{figure}

\begin{figure}[h]
\epsfxsize=7cm \epsfxsize=6cm
\centerline{\epsfbox{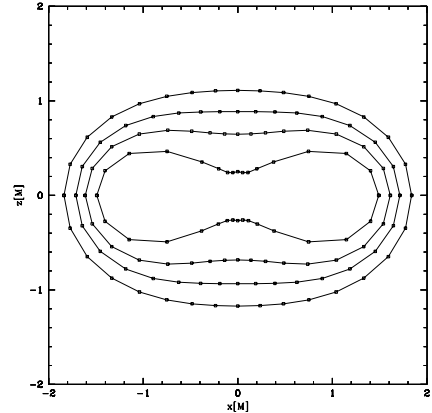}}
\caption[Level set for d=2.0M]{The level set of surface for the d=2.0M case.  Each
surface has a constant $\kappa=c_n$ at each point.  In
this case the values for $c_n$ are $c_1 = 0.14$,
$c_2 = 0.12$, $c_3 = 0.1$, and $c_4 = 0.08$.  The level
set is used to indicate the change in topology associated
with multiple surfaces.}
\label{fig:BLls}
\end{figure}

\begin{figure}[h]
\epsfxsize=7cm \epsfysize=6cm
\centerline{\epsfbox{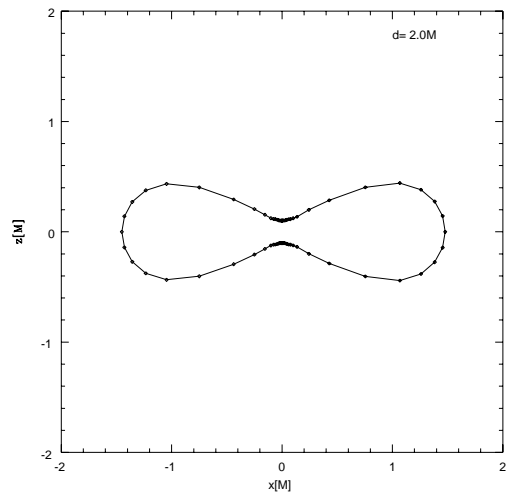}}
\caption{The single surface is about to fission into
two surfaces.}
\label{fig:tofission}
\end{figure}

As we increase the separation between the two holes to
a separation greater than the critical separation,
we can test the apparent horizon tracker in the
case of multiple apparent horizons.  We demonstrate
with a separation of $d=2.0M$.
The initial surface flows to the point of fissioning
where the topology of the surface changes from a one surface
into two surfaces.
Fig.(\ref{fig:initial}) is a plot of the initial surface
that begins the flow.  The level set found during this
flow is depicted in fig.(\ref{fig:BLls}).
Each of the surfaces in fig.(\ref{fig:BLls}) has
a constant expansion, $\kappa = c_n$ and was
used to indicate a topology change in the test surface.
The values for the expansion are $c_1 = 0.14$,
$c_2 = 0.12$, $c_3 = 0.1$, and $c_4 = 0.08$.
The last single surface just before the topology
change is not a surface in the level set; it is plotted in
fig.(\ref{fig:tofission}).  At this point the tracker
begins to flow two surfaces.

In contrast to a separation of $d=1.5M$
where there is no fission, here as fissioning
becomes imminent, the $\kappa$ begins to increase.
Fig.(\ref{fig:absy2.0}) is a plot of the
absolute value of the maximum across the surface of the expansion, $\kappa$,
versus $\lambda$ up to the point of
fission.  The increase in the expansion
is one of the signals of imminent fission.
As the algorithm tries to find a surface with
$\kappa = 0$ everywhere, it is driven into two surfaces.
Once the new surfaces are found, the maximum of the expansion begins a
monotonic
decrease as in fig.(\ref{fig:absy1.5}).  
\vskip \baselineskip

\begin{figure}[h]
\epsfxsize=9cm
\centerline{\epsfbox{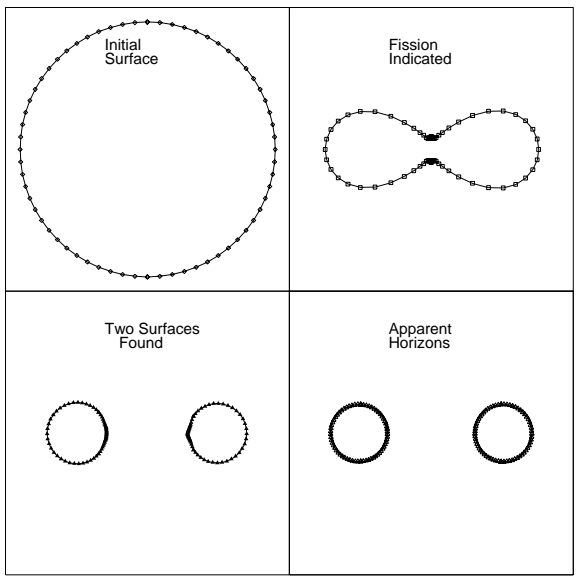}}
\vspace{0.5cm}
\caption[The flow of the surfaces in Brill-Lindquist data]
{This series of snapshots depicts the flow of an initial surface
until its fission for the binary Brill-Lindquist black holes
separated by $2M$.  The lower left plot is a first try
at determining the final two surfaces.
The cusps are due to a typical drawback associated with using points
to define the flowing surface.  The points crowd together in
regions of greater flow.  The next snapshot, on the
lower right, shows the code's automatic correction;  and shows
the apparent horizons
of the binary Brill-Lindquist data to an accuracy of $10^{-4}$.  }
\label{fig:bld1}
\end{figure}

\begin{figure}[h]
\epsfxsize=7cm
\centerline{\epsfbox{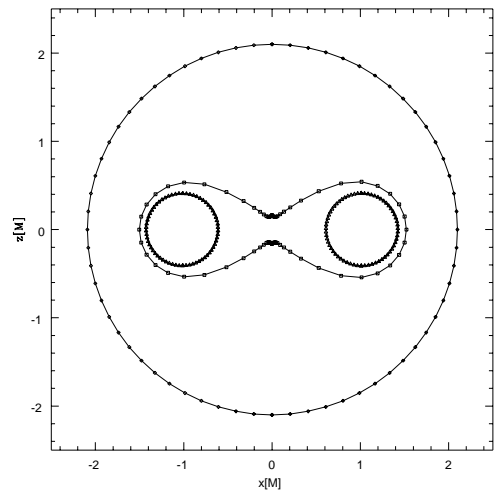}}
\vspace{0.5cm}
\caption[The flow of the surface in Brill-Lindquist data plotted together.]
{The series of pictures shown in fig.(\ref{fig:bld1})
are placed in one plot.  The outer surface is
the initial guess, the ``peanut surface" is the surface that is found indicating the need
to search for two surfaces,
and the inner approximate spheres result from locating the apparent horizons
for Brill-Lindquist data.}
\label{fig:bld2}
\end{figure}

The exaggerated peanut shape
in fig.(\ref{fig:bld1}) and fig.(\ref{fig:bld2}) is taken for the same
$\lambda$-value as  fig.(\ref{fig:tofission}).

Once the fissioning is detected by the code,
it automatically begins flowing two new surfaces
of the same resolution as the parent surface.
The series of snapshots shown in fig.(\ref{fig:bld1}) and fig.(\ref{fig:bld2})
is a subset of
the set of surfaces found by the apparent horizon tracker as it
follows the fission of the trial surface into two surfaces.
The tracker starts with a spherical starting surface that deforms
along the gradient field.
As the points defining the surface flow,
the distance between the points can become
too small for the finite difference scheme at that resolution.
Redistribution of the points on the surface is taken
care of automatically by updating the center and radius.

\section{Apparent Horizons in a Grazing Collision}
\label{sec:graze}
As stated in the Introduction, one of the main motivators
of this work is to have an apparent horizon finder that
can locate disjoint horizons during the evolution.  This entails
1) finding the horizons without a good initial guess, and 2) detecting
the topology change from two disjoint horizons to one horizon. 
To demonstrate the level flow's ability to carry out 1) and 2),
we report the results of apparent horizon tracking in the particular
case of a short evolution of
two spinning, Kerr-Schild black holes using the Binary Black
Hole Grand Challenge Alliance Cauchy code \cite{BBH}.
A future paper \cite{Grazing} will discuss the details of 
the evolution.

The evolution is free, {\it i.e.} 
the momentum and Hamiltonian constraint equations
are only used as checks during the evolution.
Since we cannot hold infinity on the grid in this
formalism, we must specify outer boundary conditions
for the dynamic variables, $g_{ij}$ and $K_{ij}$.  For the following work,
we specify analytic outer boundary conditions with blending \cite{Gomez}
between the analytic and numeric regions.

To specify initial data for two spinning,
boosted black holes we use superposed Kerr-Schild black holes.
We chose a Kerr-Schild metric [\ref{Misner}] for two reasons:
1) The metric
is well defined at the event horizon,
and 2) The metric is Lorentz form-invariant
in a simple sense, under boosts ($v>0$).
The superposed data are constructed in an approximate manner
by a conformal method based on the superposition of two isolated, 
boosted Kerr black holes. 

The initial data follows
from Matzner {\it et al.}~\cite{Matzner} and was first 
implemented by Correll ~\cite{Correll}.
The data is the superposition of two, isolated, boosted Kerr-Black holes
with individual metrics given by eqn.(\ref{eqn:KSmetric}). The
resulting superposed metric is:
\begin{equation}
\hat{g}_{ij} = {}_{(1)}g_{ij} + {}_{(2)}g_{ij} - \eta_{ij}
\label{eqn:KS2}
\end{equation}
with the  \hspace{0.1cm} $\hat{}$ \hspace{0.1cm}
symbol indicating a quantity conformally related
to the physical metric, $g_{ij} = \phi^4 \hat{g}_{ij}$.
\begin{eqnarray} \label{eqn:KS}
{}_{(1)}g_{ij}& =& \eta_{ij} + {}_{(1)}H(r_1){}_{(1)}l_i{}_{(1)}l_j \mbox{\hspace{0.5cm} and}  \\
{}_{(2)}g_{ij}& =& \eta_{ij} + {}_{(2)}H(r_2){}_{(2)}l_i{}_{(2)}l_j
\end{eqnarray}
are the the isolated Kerr-Schild metric forms
with $l_i$ and $H$ corresponding to the single black holes.
The two holes have comparable masses,
$M_1 \sim M_2$, coordinate separation $r_{12}$, and velocities
{\bf $v_1$} and {\bf $v_2$} assigned to them.
For the argument of $H$ and $l_j$, we use
\begin{eqnarray}
{r_1}^2 &=& (x-x_1)^i(x-x_1)^j \delta_{ij} \hspace{.2cm} \mbox{  and} \\
{r_2}^2 &=& (x-x_2)^i(x-x_2)^j \delta_{ij}
\end{eqnarray}
with ${x_1}^i$ and ${x_2}^j$ the coordinate positions of the holes
on the initial slice.

The extrinsic curvatures of the two isolated black holes
are added to obtain a trial ${\hat{K}}_{ab}$ for the binary black
hole system given as:
\begin{equation}
{}_{(0)}{\hat{K}}_{ij} = {}_{(1)}{\hat{K}}_{ij} + {}_{(2)}{\hat{K}}_{ij}.
\label{eqn:K0}
\end{equation}
The subscript $0$ indicates that this is an approximation to
the true extrinsic curvature of the binary black hole
spacetime.  ${}_{(1)}{\hat{K}}_{ij}$ and ${}_{(2)}{\hat{K}}_{ij}$
are the individual extrinsic curvatures associated with the
isolated Kerr-Schild metric and their indices are raised and lowered by
their individual metrics, eqn.(\ref{eqn:KS}).

\begin{figure}[h]
\epsfxsize=6cm
\centerline{\epsfbox{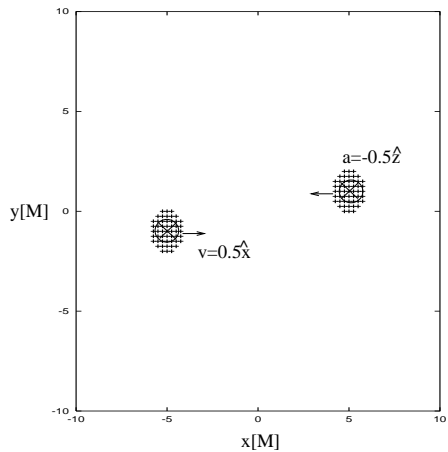}}
\vspace{0.5cm}
\caption{The configuration of the initial data for
the grazing collision.  The initial mask position is indicated
by the ``circle" centered on each hole.  The angular momentum per unit
mass and velocity of each of the holes is also represented. The total 
angular momentum (spin plus orbit) of the initial configuration is zero.}
\label{fig:KSinit}
\end{figure}

For the data we describe here (holes center initially
separated by a coordinate distance exceeding $10M$ where $M$ is the mass of
one of the holes), we expect that 
an initial value solution will be $\approx 10\%$ in error
on the domain outside of the excision volume.  See further discussion
in \cite{Pedro}.
We set the lapse function to:
$$\alpha = \alpha_1 + \alpha_2 -1,$$
and the shift vector to:
$$\beta^i = \beta_1^i + \beta_2^i.$$
The run presented in this paper 
has a grid  $81 \times 81 \times 81$ in Cartesian coordinates $(x,y,z)$
with a domain of $(\pm10M,\pm10M,\pm10M)$ resulting in a spatial resolution of 
$M/4$. The data represent two black holes in a grazing collision.
The holes
are set initially at $(5M,1M,0M)$ and  $(-5M,-1M,0M)$ in Cartesian coordinates
with a boost speed of $\pm0.5\hat{x}$ toward one another
and each has an angular momentum per unit mass of $a = 0.5M$ in the $(-)$z-direction. 
Fig.(\ref{fig:KSinit})
is the initial configuration of this run; note that a naive sum of the spin 
and the orbit angular momentum yields zero for this configuration.

\begin{figure}[h]
\epsfxsize=8cm \epsfysize=10cm
\centerline{ \epsfbox{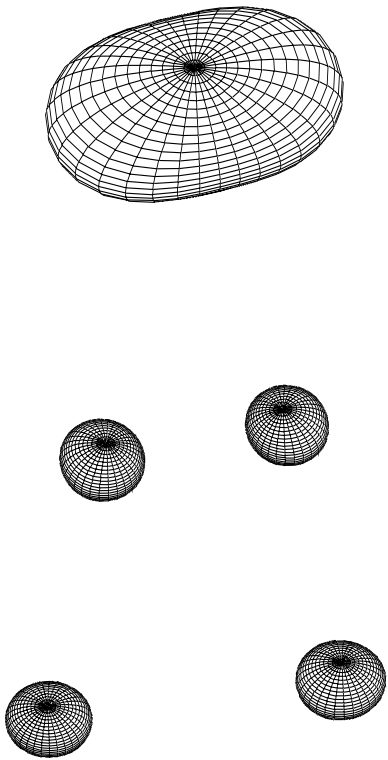} }
\vspace{0.5cm}
\caption{The apparent horizons are plotted
with the evolution time increasing up the page.
The times plotted are $0.0M$, $2.8M$, and $3.4M$.}
\label{fig:merged}
\end{figure}

We post-process the data obtained from the evolution. For the purposes
of this paper, we track the apparent horizons at three specific times
during the evolution, 
namely $t=0M$, $2.8M$, and $3.4M$.
At $0M$, the apparent horizons of the initial data are found,
at $2.8M$ two disjoint apparent horizons are found; and finally,
at $3.4M$ a single apparent horizon is found.  
For the horizons shown here, the level flow method used
a sphere of radius $8M$ to initialize each run.
Fig.~\ref{fig:merged} is a plot of the horizons with time going up the 
page.  The lowest plot is of $t=0M$, with each horizon being a sphere centered
at coordinates $(\pm 5M, \pm1M, 0M)$.  The middle plot shows the horizons at a later
time, $t=2.8M$.  Here the deviation in shape as the horizons accelerate towards
each other is seen.  The final plot at the top of fig.~\ref{fig:merged}
is the first single apparent horizon 
that envelops both black holes at $t=3.4M$.

The areas for the apparent horizons at $t=0M$ are $A = 43.6M^2$ for each hole.
At $t=2.8M$, the horizons have deviated from a spherical shape
and the  areas for each hole are $A = 44.2M^2$, giving a 
measure of the accuracy 
to which we can maintain their areas constant.
The area of the merged apparent horizon at $t=3.4M$ 
is $A_{merged} = 184M^2$.
According to the black hole area theorem
of Hawking and Ellis \cite{Hawking:Ellis}, the area
of the merged event horizon must equal at least the combined 
area of the individual
event horizons.
Although no strong statements can be made about the area
of an apparent horizon, we do find 
$A_{merged} > A_1 + A_2$ in a consistent manner.
We can further surmise that the final maximum area we could expect based
on the initial configuration should
be that of a Schwarzschild black hole of mass $2M$, giving     
an area of approximately
$201M^2$.  In some sense
the area predicted by the Schwarzschild case is an 
upper bound.  We see a $8.5\%$ deviation from that
``idealized'' case.  In view of this upper limit,
$8.5\%$ may be an indication of the greatest amount of
gravitational radiation up to the time of merger ($t=3.4M$)  
given our approximate initial data, gauge condition, and boundaries.

\vspace{1cm}

\section{Conclusion}
\label{sec:sum}

Apparent horizon location and tracking constitute an important part of
numerical evolutions of black hole spacetimes using excision techniques.
We have demonstrated a method for finding apparent horizons in situations
where the location of the apparent horizon may not be known; hence a good
initial guess for the finder may not be possible. The method we have
discussed works with generic 3-metric and extrinsic curvature data and with an
arbitrary initial starting surface. Furthermore, the method is capable of
detecting a topology change as the finder flows towards the apparent horizon.
This ability is important for situations where there are multiple apparent
horizons in the data.  This allows us to locate
apparent horizons in binary black hole evolutions without
knowing where the apparent horizons are; and
it allows us to locate the first single apparent horizon
that forms at the merger of two black holes.
The level flow method is successful at locating
the apparent horizons in generic spacetimes 
as demonstrated by the Schwarzschild and Kerr data.
It also found multiple apparent horizons starting
from a single starting surface as demonstrated with the Brill-Lindquist
data.  Most importantly, the level flow method
has been successful at identifying apparent horizons
in a binary black hole evolution involving two Kerr-Schild black holes.
Beginning with a single guess surface, two discrete apparent horizons
were found at early times, and the later single merged 
horizon was found.

One of the drawbacks of this method currently is its slow convergence 
property due to the parabolic nature of the equation solved. This, however
does not pose a problem since the level-flow method can be used in conjunction
with other methods which may be more efficient given a good initial guess. 
The level-flow method has the definite advantage of being capable of finding 
multiple surfaces in the data. It can be used to get extremely good initial
guesses for other methods that converge more quickly close to the solution.
We are currently using the level flow method in this manner
in numerical evolutions of black hole collisions.

\section{Acknowledgments}
We thank Pablo Laguna for suggesting this topic to DMS, 
Randy Correll for supplying the initial data routine used in the evolution, 
and Luis Lehner for discussions on the tracker and evolution.
This work was supported by
NSF ASC/PHY9318152, NSF PHY9800722, NSF PHY9800725 to the University of Texas
and NSF PHY9800970 and NSF PHY9800973 to the Penn State University.



\begin{references}
\bibitem{York:ADM}
\label{York:ADM}
J.~York, ``Kinematics and Dynamics in General Relativity",
{\it Sources of Gravitational Radiation}, edited by L. Smarr, Cambridge Univ. Press
(1979).

\bibitem{seidelsuen} Seidel, E. and Suen, W., Phys. Rev. Lett {\bf 69}, 1845 (1992)


\bibitem{Thornburg:1987}
\label{Thornburg:1987}
W. Unruh, quoted in J.~Thornburg, Class. Quant. Grav., {\bf 4}, 1119 (1987).

\bibitem{Anninois}
\label{Anninois}
P.~Anninos, K.~Camarda, J.~Libson, J.~Masso, E.~Seidel, W-M.~Suen,
Phys. Rev. D {\bf 58} 024003 (1998).

\bibitem{Baumgarte}
\label{Baumgarte}
T.~Baumgarte, G.~Cook, M.~Sheel, S.~Shapiro, S.~Teukolsky,
Phys. Rev. D {\bf 54} 4849-4857 (1996).


\bibitem{Gundlach}
C.~Gundlach, {\it Phys. Rev. D} {\bf 57}, 863-875 (1998). 
\label{Gundlach}


\bibitem{HCM}
\label{HCM}
M.F.~Huq, M.W.~Choptuik and
R.A.~Matzner, ``Locating Boosted Kerr and Schwarzschild Apparent Horizons'',
gr-qc/0002076, {\it Submitted to Phys Rev D}.

\bibitem{Kemball}
\label{Kemball}
A.J.~Kemball, N.T.~Bishop, ``The Numerical Determination of Apparent Horizons'',
Class. Quant. Grav. {\bf 8}, 1361 (1991).

\bibitem{Nakamura}
\label{Nakamura}
T.~Nakamura, Y.~Kojima, K.~Oohara, ``A Method of Determining Apparent
Horizons in Three-Dimensional Numerical Relativity'',
Phys. Lett. {\bf 106A}, (1984).

\bibitem{Diener}
\label{Diener}
P.~Diener, N.~Jansen, A.~Khokhlov and I.~Novikov,
``Adaptive mesh refinement approach to construction of initial data for black hole collisions,''
gr-qc/9905079.


\bibitem{Pasch}
E.~Pasch, {\it SFB 382 } Report Number {\bf 63} (1997).
\label{Pasch}

\bibitem{Thornburg}
\label{Thornburg}
J.~Thornburg, ``Finding apparent horizons in numerical relativity,''
Phys. Rev. {\bf D54}, 4899 (1996).


\bibitem{Huq}
\label{Huq}
M.F.~Huq,
``Apparent Horizons in Numerical Spacetimes,''
PhD Dissertation, University of Texas, (1996).

\bibitem{Tod}
K.P.~Tod, {\it Clas. Quant. Grav.} {\bf 8}, L115-L118 (1991).
\label{Tod}

\bibitem{Bernstein}
D.~Bernstein, {\it unpublished notes} (1993).
\label{Bernstein}


\bibitem{Grayson}
\label{Grayson}
M.~Grayson, {\it The Heat Equation Shrinks Embedded Plane Curves to
Round Points}, J. Diff. Geom., {\bf 26},  285 (1987).


\bibitem{Bruegmann}
B.~Bruegmann, {\it Int. J. Mod. Phys. D} {\bf 8} (1999) 85.
\label{Bruegmann}


\bibitem{Sethian}
S.~Osher, J.~Sethian, {\it J. Comp. Phys.} {\bf 79}, 12-49 (1988).
\label{Sethian}


\bibitem{Teukolsky}
\label{Teukolsky}
S.A.~Teukolsky, {\it Phys.Rev. D.}. {\bf 61} 087501 (2000).

\bibitem{Misner}
\label{Misner}
C.~Misner, K.~Thorne, J.~Wheeler, {\it Gravitation},
W.H.~Freeman and Co., New York, (1970).


\bibitem{Brill}
\label{Brill}
D.R.~Brill, R.W.~Lindquist, Phys. Rev. {\bf 131}, 471 (1963).

\bibitem{Cadez}
A.~Cadez, {\it Ann. of Phys.} {\bf 83} (1974) 449-457.
\label{Cadez}


\bibitem{Alcubierre}
\label{Alcubierre}
M.~Alcubierre, S.~Brandt, B.~Bruegmann, C.~Gundlach, J.~Mosso, E.~Seidel and P.~Walker,
``Test beds and applications for apparent horizon finders in numerical
                  relativity,'' gr-qc/9809004.


\bibitem{BBH}
\label{BBH}
Binary Black Hole Grand Challenge Alliance, National Science Foundation.
\begin{verbatim} http://www.npac.syr.edu/projects/bh/ \end{verbatim}             


\bibitem{Grazing}
\label{Grazing}
R.~Correll {\it et al.}, in preparation.

\bibitem{Gomez}
\label{Gomez}
R.~Gomez, in Proceedings of ``The Grand Challenge
Alliance Fall Meeting," Los Alamos, (1997).



\bibitem{Matzner}
R.~Matzner, M.~Huq, D.~Shoemaker, {\it Phs. Rev. D} {\bf 59}, 024015 (1999).
\label{Matzner}


\bibitem{Correll}
\label{Correll}
R.~Correll,
``Numerical evolution of Binary Black Hole Spacetimes,''
PhD Dissertation, University of Texas, (1998).

\bibitem{Pedro}
\label{Pedro}
P.~Marronetti {\it et al.}, {\it Phs. Rev. D} in press (2000).


\bibitem{Hawking:Ellis}
\label{Hawking:Ellis}
S.~Hawking, G.~Ellis, {\it The Large Scale Structure of Space-Time}
Cambridge University Press, Cambridge (1973).



\end{references}
\end{document}